# Title

Continuous wave laser thermal restoration of oxidized lead-based pigments in mural paintings

# Credits

# Keywords

# Authors

Théa de Seauve[a]*, Vincent Detalle[b] Alexandre Semerok[c], Sébastien Aze[d], Olivier Grauby[e], Sophie Bosonnet[f], Kevin Ginestar[f], Jean-Marc Vallet[a]

[a] Centre Interdisciplinaire de Conservation et Restauration du Patrimoine (CICRP), 21 rue Guibal, 13003 Marseille, France

[b] C2RMF, Palais du Louvre – Porte des Lions, 14, quai François Mitterrand, 75001 Paris, France

[c] Université Paris-Saclay, CEA, Service d'Études Analytiques et de Réactivité des Surfaces, 91191, Gif-sur-Yvette, France

[d] Sinopia, 1820 Chemin des Tuilières, 13290 Aix-en-Provence, France

[e] Aix-Marseille Université, CNRS – UMR 7325 CINaM, campus de Luminy, Case 913, 13288 Marseille Cedex 9, France

[f] Université Paris-Saclay, CEA, Service de la Corrosion et du Comportement des Matériaux dans leur Environnement, 91191, Gif-sur-Yvette, France

* Corresponding author: thea.de-seauve@cicrp.fr

# Abstract

Red lead and lead white are some of the most ancient and common pigments in mural paintings. However, they tend to blacken with time due to their oxidation to plattnerite ($\beta$-$PbO_2$). The possibility to induce the reconversion reactions by CW laser heating is hereby discussed. A thermodynamic study by TGA showed that direct cerussite or hydrocerussite formation from plattnerite are not suitable reconversion routes, which was confirmed by laser irradiation trials under $CO_2$ and $CO_2$/$H_2O$ fluxes. Minium ($Pb_3O_4$) and subsequent massicot ($\beta$-PbO) formation from plattnerite were achieved (confirmed by SEM-EDS, XRD and micro-Raman) under $Ar^+$, 810 nm diode and Nd:YAG lasers. The latter appears to be the most suited for restauration purposes, given the broad minium reconversion irradiance range. This is confirmed by successful trials on macroscopic areas of naturally darkened red lead containing samples.

# Highlights

- Thermal decomposition of plattnerite in $CO_2$ at 100 °C min$^{-1}$ leads to the formation of minium (490 °C) and massicot (640 °C)
- Direct formation of cerussite or hydrocerussite from plattnerite is not a suitable reconversion route
- Top-hat Nd:YAG irradiation is suitable to achieve minium and massicot reconversion for restauration purposes

# Intro

Of the many pigments that were used in easel as well as in mural paintings, lead-based pigments – particularly red lead and lead white – are definitely the earliest and most widespread [1], [2]. However, when used in mural paintings, they tends to blacken with time, which has been known for a long time [2]–[4], causing a visual imbalance that can ruin the aesthetic of the artworks [5], [6]. Oxidation to plattnerite seems to be in cause [6]–[17] in numerous murals [17]–[29]. Experimental lead white restoration treatments involving chemical reagents to achieve the reconversion[1] have been used [30], [31], but they put the pictorial layers supports at risks as recent research shows [15]–[18]. Besides, continuous wave Nd:YAG and 810 nm diode lasers were proved to be suitable tools for the thermal reconversion of blackened red lead [32], [33]. If the same technique could be applied on several other blackened pigments, it might lead to the development of a new restoration portable tool, in the same way Q-switched Nd:YAG laser is currently used for building stones cleaning [34]–[36]. Numerous advantages are associated with the use of a laser: it is intuitive and easy-to-configure, relatively cheap, and the direct control over the treated area and the use of an optical fibre make it a portable tool. Here, we address the possibility of reconverting blackened red lead and lead white on mural paintings by the use of continuous wave lasers.

Besides minium ($Pb_3O_4$), red lead pigment may contain lead monoxide (α-PbO, litharge and β-PbO, massicot), which is less stable, as an impurity resulting from the pigment manufacture process [5]. Due to the high colouring strength of minium, such impurities have little influence on the chromatic properties of the pigment [1]. Aze *et al.* wrote a review of red lead alteration in artwork, along with ascribed darkening mechanisms and aggravating conditions [5]. Red lead darkening originates from the transformation of minium into plattnerite (β-$PbO_2$) [12]–[14]. Such an alteration process has often been reported on fresco-like mural paintings [27]–[29]. No restoration method has been developed, most likely due to the complexity of the lead-oxygen system, which includes numerous crystalline varieties. Minium then massicot (β-PbO), as well as other intermediate products are formed from plattnerite when increasing the temperature at temperatures summarized in table S-2 in the SI section [37]–[45]. Thermal reduction of plattnerite under a Raman laser beam, which complicated the analysis, was reported by Burgio *et al.* [46]. Aze *et al.* successfully profited from this phenomenon to achieve the thermal reconversion of blackened minium by continuous wave (CW) Ar$^+$ (514 nm) and Nd:YAG (1064 nm) lasers irradiation [32].

---

[1] *Reconversion* is hereby used in the sense of a chemical reaction (conversion) taking place backward. Here the conversion is the oxidation to plattnerite; the reconversion is thus the formation of cerussite or hydrocerussite (white) or minium (red) from plattnerite (black) [6].

Lead white's main component is hydrocerussite ($2PbCO_3 \cdot Pb(OH)_2$), but cerussite ($PbCO_3$) can be present as well, both being in chemical equilibrium [7], [47], [48]. Depending on the "stack process" [49], [50] parameters, the hydrocerussite/cerussite ratio can range from 1:2 to 3:1 [51], [52]. Lead white was used alone, or as an admixture together with calcite, gypsum, kaolin or baryte, in a variety of media such as linseed oil, gum Arabic, egg yolk, egg tempera and animal glue [2]. The blackening of lead white in mural painting is frequent and usually ascribed to its oxidation to plattnerite, which was sparingly used as a black pigment [53]. Thermal decomposition of hydrocerussite and cerussite leading to plattnerite was already studied by various authors (tables S-3 and S-4 in the SI section summarize these studies). Tables S-5 and S-6 in the supplementary information (SI) present an exhaustive review of lead white oxidation in mural paintings [8], [9], [15], [18]–[26], [54]–[59] and in art objects [60]–[64]. Kotulanová *et al.* showed that in aqueous solution, the oxidation of lead white to plattnerite is only observed in presence of strong oxidizing agents [7]. Petushkova and Lyalikova found that $H_2O_2$-releasing bacteria, which developed in the animal glue-based binder, could be responsible for lead white oxidation [8]. Giovannoni *et al.* showed that the alkaline lime environment of *a fresco* mural painting lowers the Pb(IV)/Pb(II) redox potential sufficiently for the plattnerite formation to be caused by weak oxidizing agents such as hydrogen peroxide in high humidity conditions [15], which is in line with the Pourbaix diagram of lead [65]. Bernard *et al.* found that at pH = 8.4, the potential required to oxide hydrocerussite to plattnerite is 0.7 V [64]; incidentally, oxidation of hydrocerussite to plattnerite at pH 8.4 in chlorinated water was reported [66]. Vagnini *et al.* found evidences of plattnerite formation in *a secco* murals, which made them question the role of chlorine [9]. And later, they showed that the alkalinity of a fresco murals is not sufficient for the oxidation to be spontaneous and thus that it requires an oxidizing agent [10]. In a nutshell, plattnerite formation from lead white seems to happen in the presence of oxidizing agents, the required strength of which being determined by the alkalinity of the painting. Besides, the composition of lead white seem to play an important role, as cerussite and hydrocerussite face different alteration patterns, the latter being more reactive [7], [10], [67]. Few methods of restoration of blackened lead white have been tested [15], [18], [25], [30], [31], but they involve chemical treatments which put the support at risks [15], [18].

Our assessment of the potentiality of using CW lasers (Nd:YAG, $Ar^+$ and diode) to restore darkened red lead and lead white-containing mural paintings is hereby reported. It is grounded on a thermodynamic study of the reconversion reaction by thermogravimetric analysis (TGA) and on laser irradiation trials monitored by infrared (IR) thermography. We assessed the effectiveness of the treatment on pure plattnerite pellets, and naturally darkened mural painting samples coming from experimental ageing walls. We studied the irradiation conditions influences (laser wavelength, beam profile, irradiance, time) on the reconverted layers compositions, homogeneities and thicknesses. The latter were characterized by spectrocolorimetry, scanning electron microscopies (SEM), X-ray diffraction (XRD) and micro-Raman spectroscopy. We present here the results and discuss them within the framework of the development of a restoration tool.

# Materials and methods

This section describes our experimental approach in the main lines; experimental details and procedures can be found in the SI section.

Study of the reconversion reactions. The first part of our study aimed at determining the parameters of the reconversion reactions following equations 1, 2 and 3. Their Gibbs free energies of reaction ΔG were first calculated in order to determine in which temperature range they are thermodynamically possible. We then had to examine the thermal decomposition of plattnerite, which has been studied in air by thermogravimetric analysis (TGA) [37]–[45]. Table S-2 in the SI section summarizes the plattnerite → minium and minium → massicot reaction temperatures that were measured by these authors. On the other hand, the thermal decomposition of plattnerite in a $CO_2$ or in a $CO_2/H_2O$ atmosphere was never studied. We thus conducted this study by TGA with a $CO_2$ flux and the reactions products were characterized by means of µ-DRX. Note that, conversely, the thermal decomposition of hydrocerussite and cerussite [41], [68]–[71], sometimes in a $CO_2$ atmosphere [72]–[74] have already been studied (tables S-3 and S-4 in the SI section).

Equation 1: envisaged reaction for the reconversion of plattnerite to minium

$$3\ PbO_2 \rightarrow Pb_3O_4 + O_2$$

Equation 2: envisaged reaction for the reconversion of plattnerite to cerussite

$$PbO_2 + CO_2 \rightarrow PbCO_3 + \frac{1}{2} O_2$$

Equation 3: envisaged reaction for the reconversion of plattnerite to hydrocerussite

$$3\ PbO_2 + 2\ CO_2 + H_2O \rightarrow 2PbCO_3 \cdot Pb(OH)_2 + \frac{3}{2} O_2$$

Laser heating modelling. We then attempted to model the thermal response of plattnerite when irradiated by a CW Nd:YAG laser. We first determined some optical and thermal parameters. Its specific heat capacity, which is temperature-dependent, was measured by differential scanning calorimetry (DSC) (figure S-2 in the SI section). Its reflectance at 1064 nm was measured with a UV-Visible-NIR spectrophotometer equipped with an integrating sphere (Figure S-3 in the SI section) Other parameters were extrapolated, as explained in [75] We then used a previously developed 3D analytical model of the laser heating of materials absorbing energy from an incident laser beam. The experimental results were properly modelled in the first times of irradiation. However, after a few seconds the model does not reproduce the laser heating, due to the lack of consideration of convection and radiation phenomena [75], which is why these results will not be further discussed here.

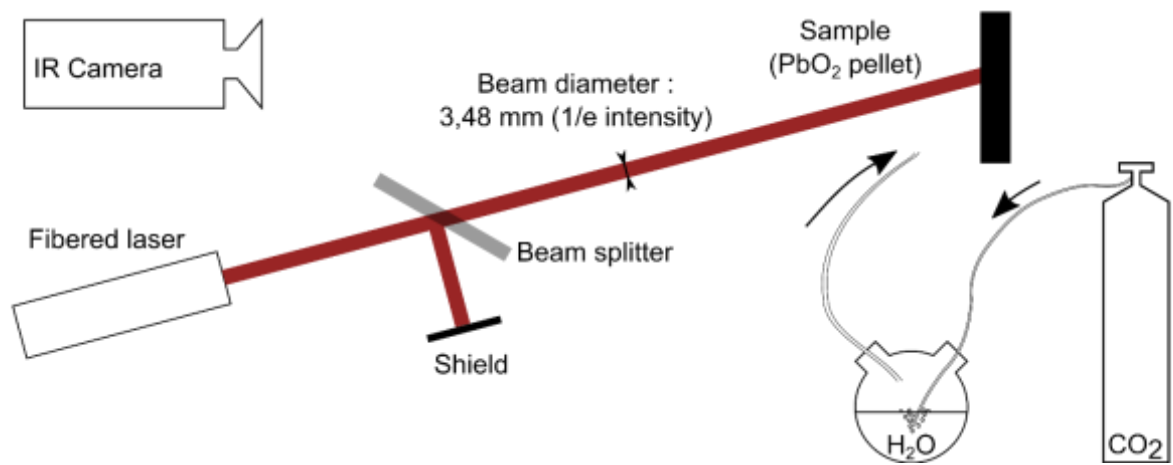

Figure 1: laser reconversion and IR thermography experimental setup

<u>Laboratory laser reconversion trials.</u> Lastly, laser irradiation trials were conducted on pure plattnerite pellets as well as on naturally aged red lead-containing paint samples. Figure 1 describes the experimental setup we used for the laser irradiation trials: the laser beam – and the gas flux ($CO_2$ or $CO_2/H_2O$) in the context of lead white reconversion – were directed toward the sample. The goal being to achieve a thermal treatment and not an ablation, the use of continuous wave lasers was preferred to that of Q-switched lasers. An IR camera was used in order to monitor the surface temperature evolution during laser irradiation. In order to do so, the emissivity of plattnerite in the IR camera detection range was first determined by heating a pellet and measuring its temperature with a thermocouple. The influence of the laser irradiance, irradiation time, wavelength and beam profile on the reconversion were assessed. Reconverted samples areas were characterized by spectrocolorimetry, SEM-EDS and XRD. SEM, Optical microscopy and Raman were also used in order to study reconverted samples cross sections. Note that Raman analysis of plattnerite is complicated by its sensitivity under the laser beam, which Burgio *et al.* and more recently Costantini *et al.* managed to assess [46], [76].

## Results and discussion

Table 1: reconversion reactions characteristics.

| Reaction | Spontaneous T range | Type |
|---|---|---|
| $PbO_2 \rightarrow Pb_3O_4$ | Above 280 °C | Endothermic |
| $Pb_3O_4 \rightarrow PbO$ | Above 420 °C | Endothermic |
| $PbO_2 \rightarrow PbCO_3$ | Below 320 °C | Exothermic |

<u>Study of the reconversion reactions.</u> Figures S-4, S-5 and S-6 in the SI section show the calculated Gibbs free energies ΔG and enthalpy ΔH of the envisaged reconversion reactions. No data were available for the plattnerite to hydrocerussite reaction. We determined the temperature thresholds at which the ΔG changes sign. The reactions are thermodynamically feasible when ΔG < 0, thus in the temperature ranges summarized in table 1. Moreover, the sign of ΔH indicates if the reactions are endothermic or exothermic in the considered temperature range. The thermal decomposition of plattnerite in air was already studied by TGA, and the compounds were identified. The plattnerite to minium reaction occurs at 375 °C [38] and the minium to massicot reaction occurs at 552 °C [41] (see table S-2 in the SI section for other results from various authors). These results are in line with the temperature threshold that we determined. We found that annealing plattnerite at 455 °C during 6 h yields pure minium (verified by XRD, as figure S-7 and table S-7 in the SI section show). In order to see if it is possible to form $PbCO_3$ from plattnerite, we studied its thermal decomposition in $CO_2$ and characterized the reaction products by µ-XRD at various temperatures (diagrams displayed in figure S-8 in the SI section). Figure 2 displays the resulting TGA curve and identification results. It shows first a small mass loss attributable to water desorption then a mass intake until *ca.* 200 °C, followed by a slight mass loss until *ca.* 400 °C. No chemical or structural change is detected in this temperature range by µ-XRD. Two major reactions then happen at 490 °C and 640 °C respectively (local minimum value of the first derivative of the TG signal). These correspond to mass losses of 4.36 % and 6.47 % respectively (local maximum value of the first derivative of the TG signal, 0.01 % precision). Measured mass losses are consistent with the theoretical values that are expected for minium (4.46 %) and massicot (6.69 %) that have been identified by µ-XRD. No cerussite has been formed in these conditions. Basing ourselves on the calculated Gibbs free energy of reaction (table 1), we let plattnerite in $CO_2$ during 50 h at 250 °C, using a slower temperature ramp (5 °C min$^{-1}$). No mass intake or loss happened and the resulting product has been identified as plattnerite. The carbonatation of plattnerite in dry $CO_2$ is probably kinetically so disfavoured that it is not a viable route to achieve the reconversion.

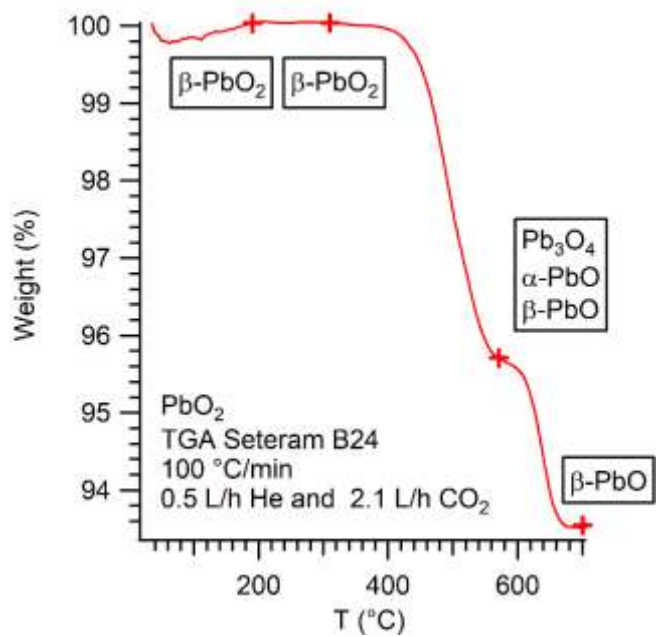

Figure 2: TGA curve of the thermal decomposition of plattnerite in $CO_2$. Insets indicate the products that were identified by µ-XRD.

IR Thermography. In order to monitor the temperature, we needed plattnerite emissivity in the measurement range of the IR camera, [7.5; 13] µm. We used the value measured by Schmidt : a 0.3 "IR emissivity" on lead dioxide coatings designed for solar panels [77]. Another source indicates a 0.28 emissivity at 258 °C in the [2.5; 50] µm range [78]. We were then able to monitor the temperature evolution during laser irradiation with the IR camera (figure 1). Figure 3a displays the evolution of radial temperature during laser irradiation, using a CW Gaussian shaped 1080 nm diode laser. At 1 s, the profile is quasi-Gaussian, with a ΔT dropping at 1/e of the maximum at a 1.8 mm radial distance which is close to the laser beam 1/e radius. As the irradiation goes on, the temperature variation profile gets further from a Gaussian profile, with an increasing radius at 1/e of maximum ΔT (figure S-9 in the SI section). This can be explained by lateral thermal diffusion happening at long irradiation times. Figure 3b shows the maximum temperature measured near the centre of the pellet. When no reconversion is observed it reaches a plateau which is higher as the laser irradiance increases. Energy can be lost by convection, radiation and diffusion, leading to the saturation that is observed. However, the broadening of the temperature profile that is observed on figure 3a indicates that diffusion is probably the most important phenomenon. Figure 3c shows the plateau temperature rising linearly with the laser irradiance ($r^2 = 0.9961$ for a linear fit). This is expected and confirms that neither the emissivity nor other optical and thermal parameters of plattnerite vary much over this temperature range. Above 152 kW m$^{-2}$, which corresponds to a plateau temperature of 375 °C, plattnerite is expected to decompose into minium, according to the reaction temperature that was measured by Clark *et al.* [38].

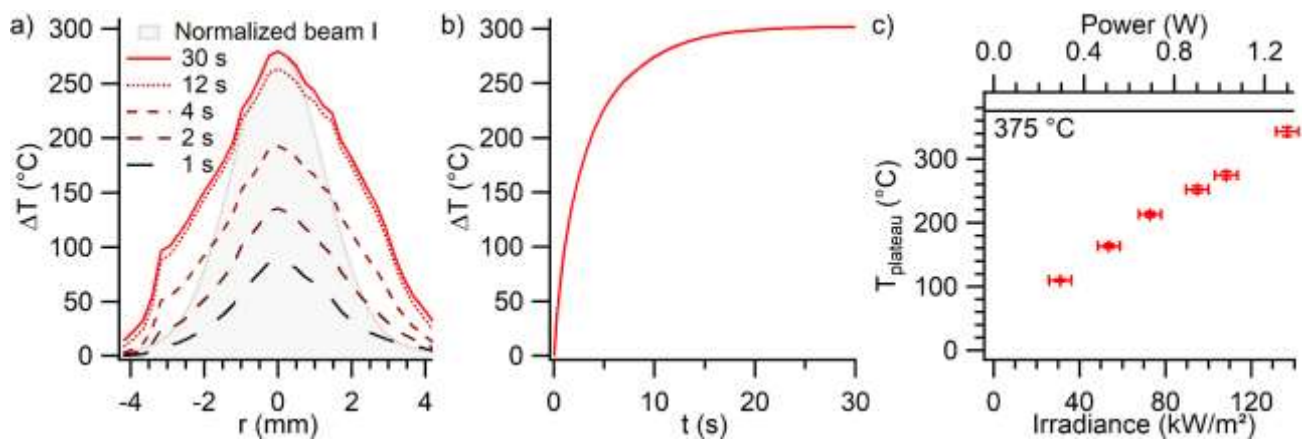

Figure 3a) Radial surface temperature rise of a 317 µm thick plattnerite pellet exposed by the Gaussian shaped 1080 nm laser ($r_{1/e}$ = 1.74 mm) at a 137 kW m$^2$ irradiance at various irradiation times. The best Gaussian fit of the beam intensity is plotted in grey. b) Evolution of the maximum measured surface temperature in the same conditions as figure 3a. c) Plateau temperatures measured on a plattnerite pellet exposed at various irradiances by the Gaussian shaped 1080 nm laser.

Laser reconversion trials. The thermal reduction of plattnerite to minium that indeed occurs in the same conditions at irradiances higher than 152 kW m$^{-2}$ is clearly noticeable under the form of a red/orange spot (figure 4a). Modifications of the plattnerite pellets colours occurred between 5 and 10 seconds of irradiation, in accordance with figure 3b. No further visual or chemical difference is observed at higher irradiation times. Spectrocolorimetric measurements show an increase of the specular reflectance at wavelengths longer than 560 nm (Figure 4b). This is consistent with the optical properties of plattnerite and minium (see figure S-10 in the SI section for the standard spectra of these compounds). The corresponding L*a*b* values indicate that the irradiated area turns to red with a ΔE of 34. We found a good agreement between the experimental reconversion threshold and the threshold we predicted from IR thermography measurements and Clark *et al.*'s results (figure 3c). This can be interpreted as a confirmation that the IR emissivity we measured is plausible. Above the reconversion threshold, the temperature keeps increasing without showing no plateau and then decreases slowly. Minium absorbs less than plattnerite at the laser wavelength (figure 3b) and would thus be less heated by the laser, which can be a possible explanation to this. Other changes in the thermal property can also explain this observation. Moreover, the endothermicity of the reaction can also be part of the phenomenon. However, the fact that both plattnerite and minium can be present in these samples and that they have somewhat different IR emissivity (minium emissivity is 0.93 at 398 °C [78]) prevent us to discuss these IR thermography data further.

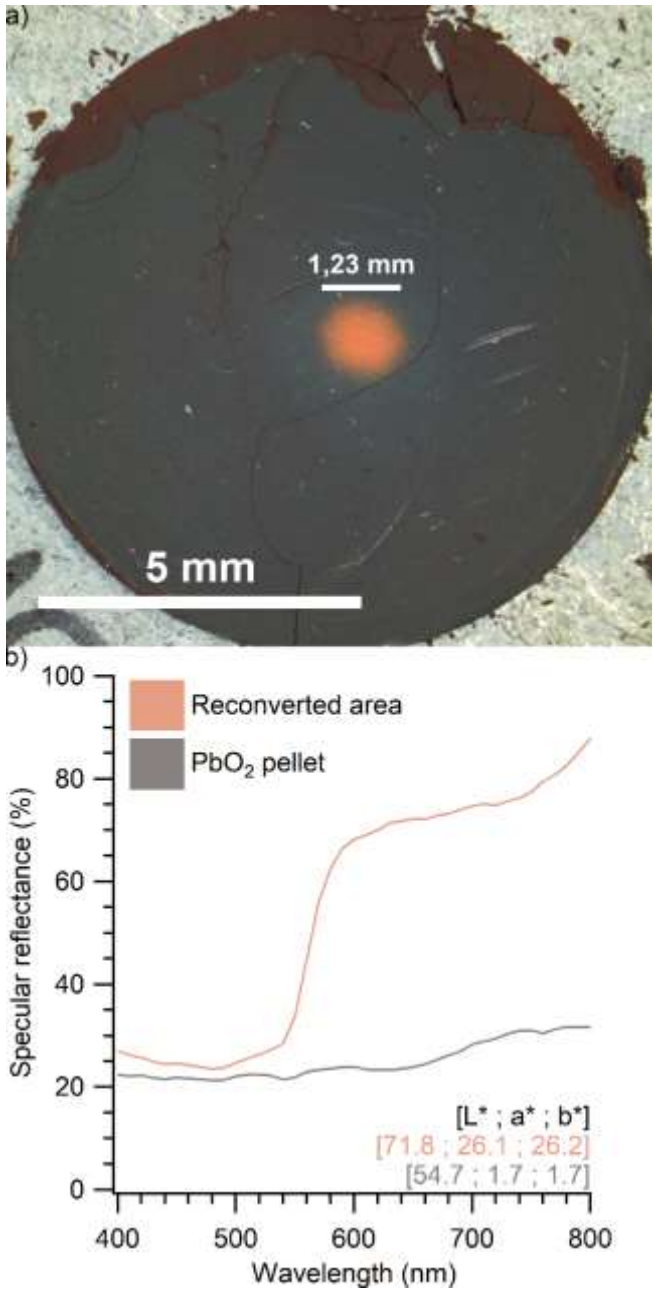

Figure 4a) Pure 440 µm-thick plattnerite pellet that has been exposed to a 3.48 mm diameter 1080 nm diode laser beam (1/e value) at a 205 kW m$^{-2}$ irradiance during 2 min. b) specular reflectance spectra of the irradiated and pristine areas of the same pellet. Colour patches have been displayed using the RGB values calculated from the L*a*b* coordinates.

Power density threshold. The above results were obtained with a Gaussian shaped laser beam. The energy being concentrated in the inner part of the beam (figure 5b), the surface temperature reached during irradiation is higher when closer to the beam centre. The diameter of the reconverted minium area on figure 4a is 1.23 mm. It was obtained at 205 kW m$^2$ with a 3.48 mm beam diameter (1/e value). According to the Gaussian distribution of power, the reconversion is thus effective at power densities higher than 181 kW m$^{-2}$.

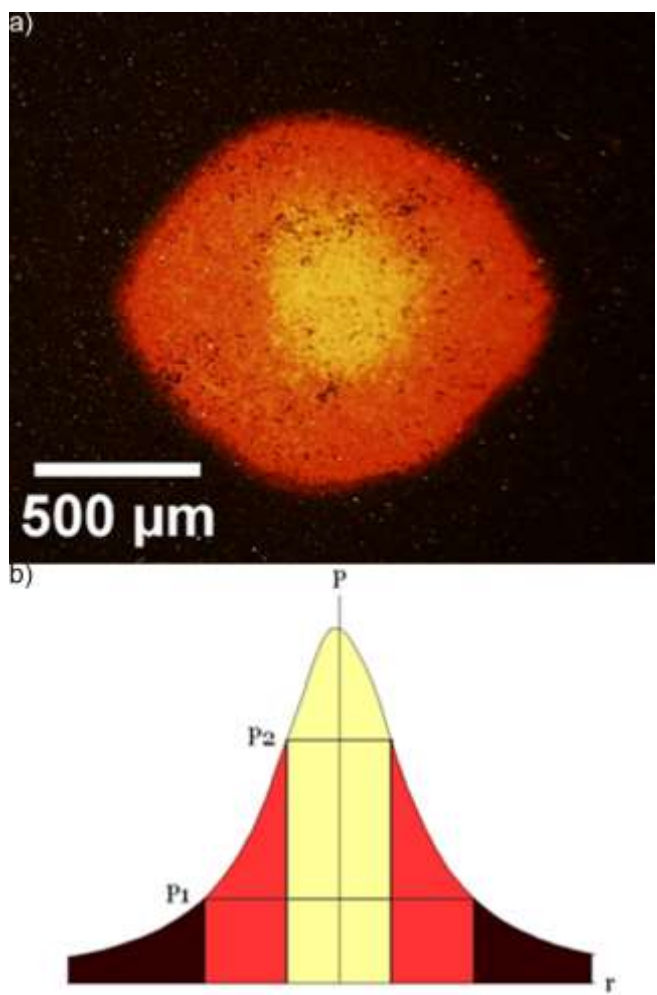

Figure 5a) Massicot (yellow) formed in the centre of a reconverted minium area (red/orange) at high $Ar^+$ laser irradiance. b) Gaussian laser beam profile which explains the phenomenon.

Massicot formation at high irradiances. Using a top-hat CW Nd:YAG laser a yellow compound, later identified as massicot β-PbO, was created at irradiances higher than 2605 kW $m^2$. This is expected, since massicot is formed from minium at 552 °C [41]. The wide range between the irradiance threshold of reconversion to minium and to massicot is likely due to the fact that minium absorbs less in the infrared range than plattnerite, and is thus heated less efficiently. This can be a problem when using a Gaussian shaped laser beam, as shown by figure 5a: the central area has been reconverted to massicot whereas a minium ring is formed around massicot, as the micro-Raman identification show (figure S-11 in the SI section). In this case, the power density is so high at the centre of the beam that massicot is formed, and high enough around the centre of the beam to form minium (figure 5b). This phenomenon can be a problem in the context of a restauration process. Therefore, the use of a top-hat laser beam profile would be better.

Table 2: minium and massicot reconversion thresholds for various lasers. Values have been calculated by using the 1/e beam radius.

| Source | Plattnerite to minium (kW $m^{-2}$) | Minium to massicot (kW $m^{-2}$) |
|---|---|---|
| CW Gaussian shaped diode (1080 nm) | Laser irradiance: 152<br>Effective power density: 181 | |
| CW top-hat Nd:YAG (1064 nm) | 109 | 2605 |
| FAP-S Pb10 profile (810 nm) | 275 | 504 |
| CW Gaussian shaped $Ar^+$ (488 nm) | 136 | 151 |

Laser wavelength influence. We used an Ar$^+$ (488 nm) and a FAP-S (810 nm) laser and conducted the same reconversion trials. We observed the same processes of minium then massicot formation, although at somewhat different irradiance, summarized in table 2. The minium reconversion threshold is higher for the Ar$^+$ than for the 1080 nm diode laser, both being Gaussian-shaped. A higher absorbance of plattnerite at 1080 nm than at 488 nm could explain this difference. The various beam profiles make it risky to compare the other reconversion thresholds. However, we can still notice that the irradiance range in which minium is formed, before it turns to massicot, is wider in the case of the Nd:YAG laser (1064 nm) than in the case of the FSP-S diode laser (810 nm), which is itself wider than in the case of the Ar$^+$ laser (480 nm). This is explained by the considerably higher absorption of shorter wavelengths by minium than higher wavelengths. In practice, a suitable Ar$^+$ laser irradiance for plattnerite reduction into pure minium was practically unreachable. A secondary transformation into massicot was subsequently achieved through excess local heating, which makes this laser unsuitable for restoration purposes.

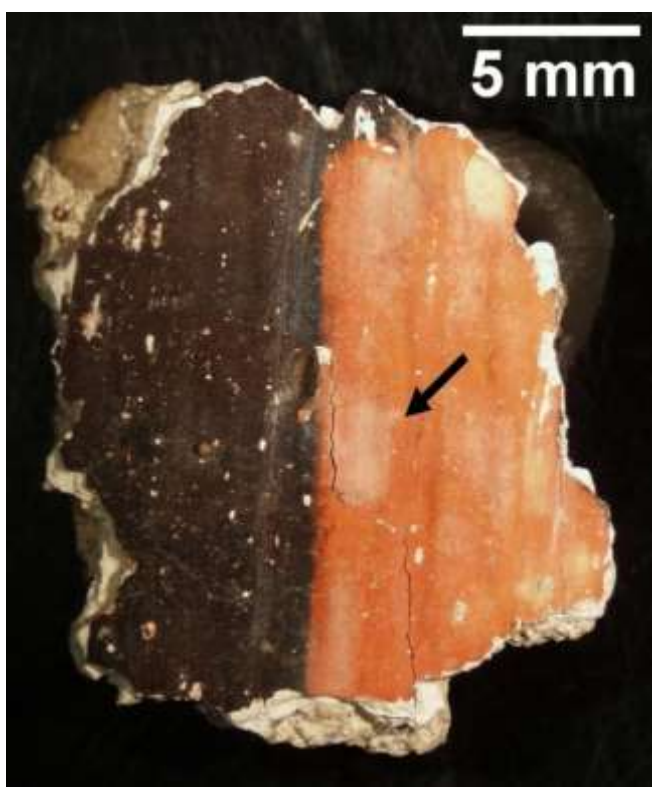

Figure 6: darkened red lead paint sample after irradiation by a CW top-hat Nd:YAG laser (red part on the right) at 1000 kW m$^{-2}$. The lighter stains designated by an arrow are due to gypsum. Cracks were present before the treatment.

Naturally aged samples. Irradiation tests on darkened red lead-containing paint samples showed the same results. However, the irradiance thresholds were notably higher than those measured for pure plattnerite pellets. It is possible that such differences originate from the higher reflectivity of naturally aged paint samples. This latter is attributed to the presence of gypsum, which appeared during ageing, but also to the brown colour of finely divided plattnerite particles. It can also be attributed to differences in the surface state of the samples, in terms of rugosity or porosity. Large-scale irradiation tests were carried out at by translating the sample under the laser beam along two directions, with a mean local irradiation time of *ca.* 5 seconds. The laser irradiance (1000 kW m$^{-2}$ calculated by using the 1/e beam radius) was selected so that all plattnerite grains were transformed into minium. Such treatment produced a quasi-uniform red area (figure 6). The presence of lighter stains arises from gypsum, which did not evolve during laser irradiation. Observations of the sample cross-section (figure S-12 and S-13 in the SI section) show that the whole darkened layer has been reverted into a 50 to 80 µm thick red homogeneous phase. Such depth efficiency may result from the heat diffusion within the alteration layer. XRD and micro-Raman analyses of this phase confirmed the presence of minium, without any lower lead oxide in the reconverted layer (figure S-14 and S-15 in the SI section).

<u>Lead white reconversion.</u> Laser trials were also made on plattnerite with a $CO_2$ or $CO_2/H_2O$ input. The same reactions were observed, but the gaseous flux had a cooling effect and the reconversion irradiance threshold were thus higher. No specie other than minium or massicot were found. This is consistent with the TGA results, which show no carbonatation of plattnerite in $CO_2$ and clearly indicates that the reactions described by equations 2 and 3 are not suitable reconversion routes. Minium [79]–[82] and PbO [83], [84] can react with $H_2O$ and $CO_2$ at the solid state to form hydrocerussite or cerussite. Note that the formation of plumbonacrite, a white compound, from minium has also been evidenced [85]. Often, the formation of cerussite or hydrocerussite from PbO (generally identified as massicot) is observed spontaneously in air at ambient temperature when the latter is the degradation product of lead white irradiated by a Q-switched Nd:YAG laser beam [86]–[91]. Hence, it could be possible to form cerussite or hydrocerussite from plattnerite *via* the intermediary formation of massicot and its subsequent exposition to ambient $CO_2$ or $CO_2/H_2O$. Studying the latter reaction is a first step toward the assessment of the possibility to reconvert darkened lead white by this route.

# Conclusion

The formation of cerussite from plattnerite in $CO_2$, an exothermic reaction spontaneous under 320 °C was not observed during thermogravimetric analysis. Only the formation of minium (at 490 °C) massicot (at 640 °C) were observed. We monitored the temperature during laser irradiation trials by IR thermography. In air, we observed an irradiance threshold under which the temperature rises to a plateau and above which reconversion to minium takes place at the expected temperature (375 °C). Plateau temperature rises linearly with irradiance, indicating that plattnerite thermal and optical parameter vary little in this temperature range. Spectrocolorimetric measurement show a reddening of the sample with a ΔE of 34, compatible with a reconversion to minium. At high irradiances, the formation of massicot is also observed. Minium and massicot reconversion thresholds depend on the laser wavelength. Nd:YAG minium reconversion range ([109; 2605] kW m$^{-2}$ in the case of a top hat laser) is wider than AR$^+$ and 810 nm diode lasers ranges. Successful trials were made on naturally darkened red led containing paint samples, moving and spending 5 s under the 1000 kW m$^{-2}$ top-hat Nd:YAG beam. XRD and micro-Raman indeed detected the presence of pure minium in reconverted layers. The reconversion of plattnerite to lead white by a 1080 nm diode laser beam under gaseous flux ($CO_2$ or $CO_2/H_2O$) did not happen, in line with TGA results.

We can conclude from these results that fibered lasers are suitable to achieve the reconversion of minium and massicot in mural paintings. Particularly, the top-hat Nd:YAG laser is best adapted for restauration purposes, given its broad minium reconversion irradiance range and its homogeneous power density distribution. It was shown to yield good results for the treatment of a large area of a naturally aged sample, showing its suitability in restauration conditions. The direct route we assessed for lead white reconversion, namely the carbonatation of plattnerite, was however unsuccessful. The formation of massicot as an intermediate product, which could later react with ambient $CO_2$ and $H_2O$ to form cerussite and hydrocerussite, is a promising reconversion route.

# Acknowledgement

The authors wish to acknowledge the French Ministry of Culture and Communication for financial support, and the PLANI group from the LILM laboratory (CEA) which supplied their experimental set up. The authors also wish to acknowledge the financial support of the French Fondation des Sciences du Patrimoine to the RECONVERT project, as well as Kamel Mouhoubi (ITHEMM, Université de Reims) for the interesting discussions and details about IR thermography and the determination of emissivity. The authors would also like to thank Philippe Delaporte (LP3) and Alain Baronnet (CINaM) for their contribution to preliminary studies. Lastly, the authors would like to thank Emmanuel André (LP3) for granting access to the Shimadzu UV-visible-NIR spectrophotometer.

# Supplementary information

## Experimental section

Pure plattnerite samples. Pure 150 to 700 µm thick and 10 mm diameter plattnerite pellets were pressed (three times at 100 kN for 10 mins) from 97 % pure commercially available powder (Alfa Aesar A12742, Lot number 10213539). XRD showed no difference between the powder and the pressed pellet.

Naturally aged red lead-containing paint samples. These were taken from experimental wall paintings [92]. Previous investigations of red lead alteration revealed the transformation of minium phase into plattnerite in the darkened areas [12]. In addition, most of the darkened samples contain gypsum, most likely as a result of the sulfation of the original calcic material.

ΔG calculations. The Gibbs free energies of reaction were calculated with the HSC Chemistry software.

TGA. A Setaram B24 thermogravimetric analyser was used in the following conditions. We degassed the crucibles at 1000 °C for a few days before use. The sample crucible is usually loaded with *ca.* 90 mg $PbO_2$ and the crucible usually weight *ca.* 325 mg. Before each trial stabilization of the system took place at 30 °C for 3 h. The crucibles are subjected to a 0.5 L min$^{-1}$ Ar flux and a 2.1 L min$^{-1}$ $CO_2$ flux. The mass change detection limit is equal to 10 µg in these conditions. A blank trial was first made with an empty crucible as a baseline for an exploratory trial until 700 °C. The samples were heated up to the desired temperature 100 °C min$^{-1}$, then spent 10 min at the plateau temperature before being cooled down to ambient temperature at 100 °C min$^{-1}$. Temperatures corresponding to assumed chemical changes were then selected and the same profile was chosen to get to these temperatures. Resulting products were characterized by µ-XRD.

µ-XRD. The XRD measurements were carried out using a device mounted on a high-gloss rotating anode, Rigaku RU-200BH, the anode being made of Cu. A reflective focusing optic, OSMIC, mainly transmits the Cu Kα radiation (λ = 1.5418 Å) and a very small part of the Kβ, the latter being absorbed completely by a Ni filter. The detector used is 2D, flat image type, model Mar345. For these measurements, the working power was 50 kV and 50 mA and the beam size was 0.5 × 0.5 mm$^2$. XRD patterns were collected from quartz capillary containing crushed samples.

DSC. A Netzsch ST409PC was used at a 2 °C min$^{-1}$ rate and with 60 and 20 mL min$^{-1}$ $N_2$ flows on the SiC heating element and on the microbalance respectively. The device was temperature- and sensitivity-calibrated with the following standards: In, Sn, Bi, Pb, Zn. Samples were weighted with a Sartorius LA230S balance (0.1 mg precision) and placed in Netzsch Al crucibles (6.239.2-64.5.01 and .02). DSC measurements were first made on an empty crucible to establish a baseline, then on a crucible with 15.1 mg of Al and a crucible with 36.65 mg of $PbO_2$. The specific heat capacity of $PbO_2$ was then calculated according to equation 4, with Cp values for Al given by Hrbek [93].

Equation 4: specific heat capacity calculation by the ASTM method with calibrated DSC curves

$$C_{p,PbO_2} = \frac{m_{Al}}{m_{PbO_2}} \times \frac{DSC_{PbO_2} - DSC_{baseline}}{DSC_{Al} - DSC_{baseline}} \times C_{p,Al}$$

Total reflectance spectroscopy. The UV-Visible-NIR spectroscopic measurements were performed on a Shimadzu UV-2600i spectrophotometer equipped with an integrating sphere (ISR-2600+) (1 nm step, 0.5 s nm$^{-1}$).

Spectrocolorimetry. A RUBY spectrocolorimeter was used to acquire the specular reflectance spectra in the [400; 800] nm of the samples placed 80 mm away from the detector. A Labsphere certified white reflectance standard (USRS-99-010 AS-01158-060, serial number: 99A10-0513-9483) was used as the white reference and a black cardboard was used as the black reference. 100 spectra of the references and of the samples were acquired with a 10 nm step at a 840 Hz rate and averaged. The software automatically calculates the colorimetric coordinates in the CIE 1964 space with a D65 illuminant. Note that this spectrocolorimeter is designed to measure colour differences between samples analysed together.

Microscopy. An Olympus SZX7 stereo microscope equipped with a UC30 camera run by the Stream Basic software was used to take pictures of the samples

XRD. A Bruker D8 Focus diffractometer has been used in the $\theta - 2\theta$ mode, with a 0.01° step and 1 s/step. The X-ray source is a Co anticathode ($\lambda = 1.789$ Å) operated in the following conditions: 40 keV / 35 mA.

Temperature measurement. During laser irradiation experiments, the temperature was measured with a FLIR SC640 IR camera equipped with a T197089 (f = 38 mm) objective, operated by the ThermaCam software and located 50 cm away from the sample. Its spectral measurement range is [7.5; 13] µm. Filmed in these conditions, a 1 cm pellet results in a 35 pixels diameter circular area. Reported accuracy is 2 °C or 2 % of measured temperature.

Gaussian-shaped 1080 nm laser experimental setup. Most of the experiment were made with a 1080 nm diode laser. A fibered Manlight ML50 CWR-TKS laser, of peak wavelength 1080 nm, was used together with a Schlumberger 4433 periodical signal generator. The laser was used either in continuous mode. A beam splitter was occasionally used to decrease the beam power, which was controlled with a Fieldmate laser power meter and a Coherent powermax PM30 and ranged from a few 0.1 W to a few W. Beam power is measured after stabilization, with an estimated accuracy of 0.05 W. Lenses were occasionally used to increase or decrease the beam radius. A Spiricon laser beam profiler was previously used to determine the Gaussian-shaped beam radius, which was 1740 ± 30 µm at 1/e of peak intensity [94]. Samples were irradiated for 1 s to 120 s.

Table S-1: main characteristics of the CW laser sources employed for the irradiation tests.

| Type | λ (nm) | Supplier | Model | Max power | Profile | 1/e beam radius (mm) |
|---|---|---|---|---|---|---|
| Ar$^+$ | 488 | Coherent | Innova 200-10 | 10 W | Gaussian | 1.45 |
| Diode | 810 | Coherent | FAP-S | 30W | | 1.86 mm at 9 cm |
| Nd:YAG | 1064 | LILM | | 60W | Top-hat | 0.66 |
| Diode | 1080 | Manlight | ML50 | 50 W | Gaussian | 1.74 |

Ar$^+$, FAP-S and Nd:YAG laser. In order to investigate the influence of the irradiation wavelength and beam profile, other CW laser were investigated. Their characteristics are summarized in table S-1. Laser power emission curves versus current were ascertained through recurrent measurements using a Powermax 500A laser power meter. Quasi-parallel, circularly symmetric laser beam were obtained using a set of optical devices, including both converging and diverging lenses, diaphragms and mirrors. Beam power density profiles were established using a Gentec Beamage CCD13 analyzer.

For the Nd:YAG laser, a quasi "top-hat" energy profile was obtained, while $Ar^+$ beam profile was measured as near-Gaussian (figure S-1).

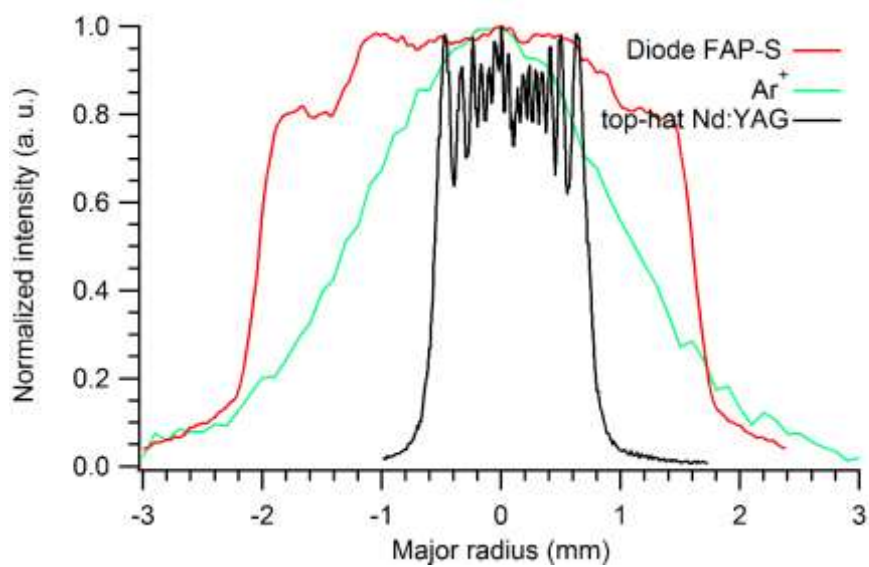

Figure S-1: radial power density profiles of the $Ar^+$, top-hat Nd:YAG and FAP-S diode laser beams.

Gaseous flux during laser irradiation. Expended carbon dioxide from the bottle, occasionally bubbled through distilled water, was directed at the sample, the latter and the nozzle spaced 5 cm apart. The estimated flux was 10 L min$^{-1}$ for $CO_2$ alone and 2 L min$^{-1}$ for $CO_2/H_2O$. Note that at ambient temperature, the vapour pressure of water is limited to 2348.6 Pa, that is 2.32 % in mole fraction [95].

Raman microspectroscopy. Micro-Raman spectra of representative areas were collected using a Renishaw InVia system, equipped with a GaAs diode laser source (785 nm). For individual particles observed under 1000× microscope, both confocal and pinhole diaphragm accessories were used so that no external signal was detected. Measurement parameters were selected so that no further photo-thermal degradation occur during spectral acquisition [46].

# Supplementary material

Table S-2: measured temperatures of thermal decomposition of plattnerite in chronological order of publication. Note that both the temperature profile [37], [44] and the initial granulometry [45] can have a critical impact, not only on the reactions temperatures, but also on the reaction product granulometry [37].

| Ref | $3\ PbO_2 \rightarrow Pb_3O_4 + O_2$ | $Pb_3O_4 \rightarrow 3\ PbO + ½\ O_2$ |
|---|---|---|
| [38] | [355; 375] °C | |
| [40] | 498 °C | |
| [43] | | [480; 640] °C |
| [39] | 450 °C | 650 °C |
| [41] | | [607; 625] °C (in $O_2$) |
| | | [512; 552] °C (in air) |
| [42] | | 535 °C |

Table S-3: measured temperatures of thermal decomposition of hydrocerussite in chronological order of publication. $x = 1, 2$ or $4$; $y = 1, 2$ or $3$; $x/y$ reduces with increasing temperature [41].

| Ref | $2PbCO_3·Pb(OH)_2 \rightarrow xPbCO_3·yPbO + H_2O$ | $xPbCO_3·yPbO \rightarrow PbO + CO_2$ |
|---|---|---|
| [68] | [235; 290] °C | [380; 505] °C |
| [74] (in $CO_2$) | [170; 300] °C | [300; 420] °C |
| [72] (in $CO_2$) | [150; 275] °C | [450; ~530] °C |
| [41] | [210; 260] °C | [320; 370] °C |

Table S-4: measured temperatures of thermal decomposition of cerussite in chronological order of publication. $x = 1, 2$ or $4$; $y = 1, 2$ or $3$; $x/y$ reduces with increasing temperature [70].

| Ref | $2PbCO_3 \rightarrow xPbCO_3·yPbO + CO_2$ | $xPbCO_3·yPbO \rightarrow 2\ PbO + CO_2$ |
|---|---|---|
| [68] | [320; 395] °C | [395; 310] °C |
| [71] | [340; 390] °C | 440 °C |
| [70] (in $CO_2$) | [204; 373] °C | 435 °C |
| [73] | [276; 350] °C | [353; 420] °C |

Table S-5 (next page): occurrences of lead white darkening on mural paintings due to plattnerite formation. The date refers to the time where the murals have been painted. The certainty relates to the attributed origin of plattnerite, (from minium or from lead white).

| Date | Site | Mural | Technique | Identification | Certainty | Reference |
|---|---|---|---|---|---|---|
| XIII[th] | San Francesco Basilica in Assisi | Cimabue's mural paintings | *a secco* | XRD, Raman | Certain | [9], [15] |
| | San Miniato al Monte, Florence, Italy | Paintings by Baldovinetti | | | | [15] |
| | Cloister of the Abbey of Monteoliveto Maggiore, Asciano, Italy | Frescoed lunette by Luca Signorelli | | | | [15] |
| XII[th]–XIII[th] | David Gareja monastery complex, Georgia | Central Churches of Bertubani, Udabno and Natlis-Mtsemeli monasteries | On gypsum plaster with *tempera* technique "in secco" | Optical microscopy, XRD, Mass spectroscopy | Certain | [19], [20] |
| XV[th] | Saint Stephen's chapel, Morter, Italy | Late Gothic wall paintings | Beeswax? | Raman | Certain | [18] |
| XV[th] | S. Maria ad Undas, Idro, Italy | Masonry altar | Organic residuals | Raman | Certain | [21] |
| XVI[th] | Church of Madonna della Difesa, Vigo di Cadore, Italy | Abside lawn painting (V12) | *Secco* | XRD, SEM-EDS, Optical microscopy | Certain | [22] |
| XVIII[th] | Franciscan monastery cloister, Saorge, France | Murals relating Saint Francis of Assisi's life | *Tempera* | XRD | Certain | [23] |
| XIV[th] | Château de Germolles, Mellecey, France | The noble floor wall paintings | Linseed oil | Naked eye observation | | [24] |
| 1580 | Castle of Strechau, Lassing, Austria | Chapel ceiling paintings | dry lime plaster, lead white and chalk, pigments in glue *tempera* | | | [25] |
| XVI[th] | The Santa Cruz Monastery, Coimbra, Portugal | Tomb of the first Portuguese King D. Afonso I | | Raman | | [26] |
| XVII[th] | Church of St John the Divine, Rostov the Great, Russia | Deesis range on the altar wall | | "X-ray examination" (XRD) | | [8] |
| IV[th] | Dominican Monastery, Ptuj, Slovenia | Gothic wall painting | Lime | Raman | Possible | [54] |
| XII[th]–XIII[th] | Holy Sepulchre Chapel, Winchester Cathedral, United Kingdom | Romanesque and Gothic Wall Paintings | *A fresco*, *a secco* | Naked eye observation, SEM/EDX | Possible | [55] |
| XII[th] | Yemrehanna Krestos Church, Lalibela, Ethiopia | Medieval wall paintings | *Fresco secco* on gypsym | Portable XRF | Uncertain (can be PbS) | [56] |
| | Santa Clara Church, Sabugueiro, Portugal | Flesh tones of the angel figures | | Raman | Uncertain | [59] |
| VIII[th] | Chapel of the Holy Cross, Müstair, Switzerland | Carolingian wall paintings, upper floor | Pure dolomitic lime (whitewashes) Organic binders (colours) | Multispectral imaging | Unlikely | [57] |
| XI[th] | Saint George Church, Kostoľany pod Tribečom, Slovakia | Early Romanesque wall paintings | | μ-XRD | Unlikely | [7] |
| XV[th] | St. Orso Priory palace, Aosta, Italy | Chapel wall paintings | | Raman | Unlikely | [58] |

Table S-6: occurrences of lead white darkening on objects due to plattnerite formation.

| Date | Object | Technique | Identification | Certainty | Reference |
|---|---|---|---|---|---|
| 1480 – 1500 | "Trionfo d'Amore", manuscript no. 143 by Sandro Botticelli (Biblioteca Classense, Ravenna, Italy) | Drawing | Raman | Certain | [60], [61] |
| XVII[th] | Bertesi's wooden bas relief (Fondazione Città di Cremona, Cremona, Italy) | "Stucco-like" white-coated wooden bas relief | Naked eye | | [63] |
| 1650 | 'Cembalo' model musical instrument by Michele Todini (Museum of Palazzo Venezia, Rome, Italy) | Pottery | Raman | Certain | [62] |
| XVIII[th] | Lead statues in the garden of the Queluz Palace, Sintra, Portugal | Weathering induced lead white | Raman, XRD | | [64] |

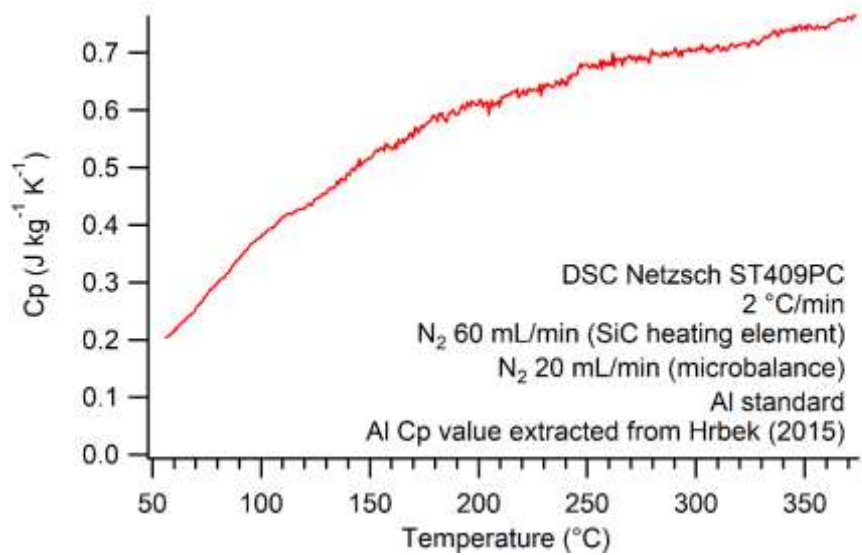

Figure S-2: specific heat capacity of plattnerite β-PbO$_2$ measured by DSC

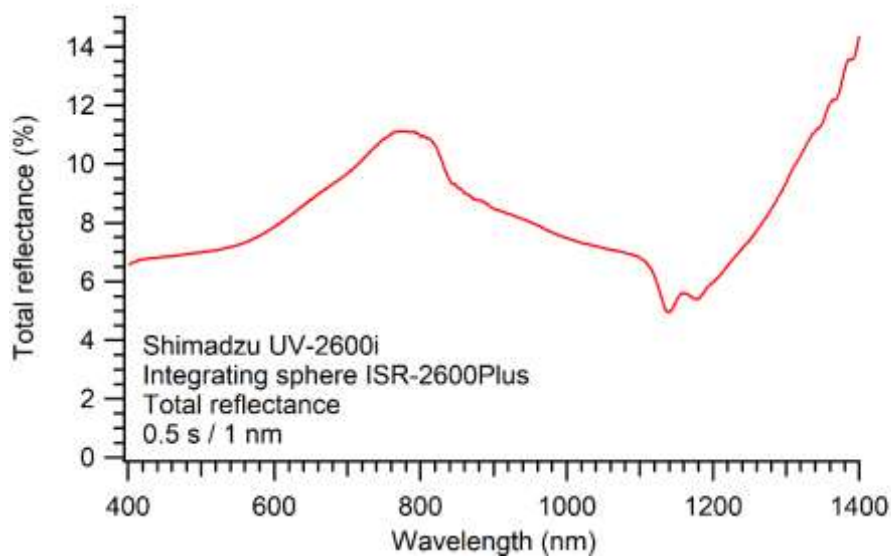

Figure S-3: total reflectance spectra (diffuse + specular) of plattnerite β-PbO$_2$

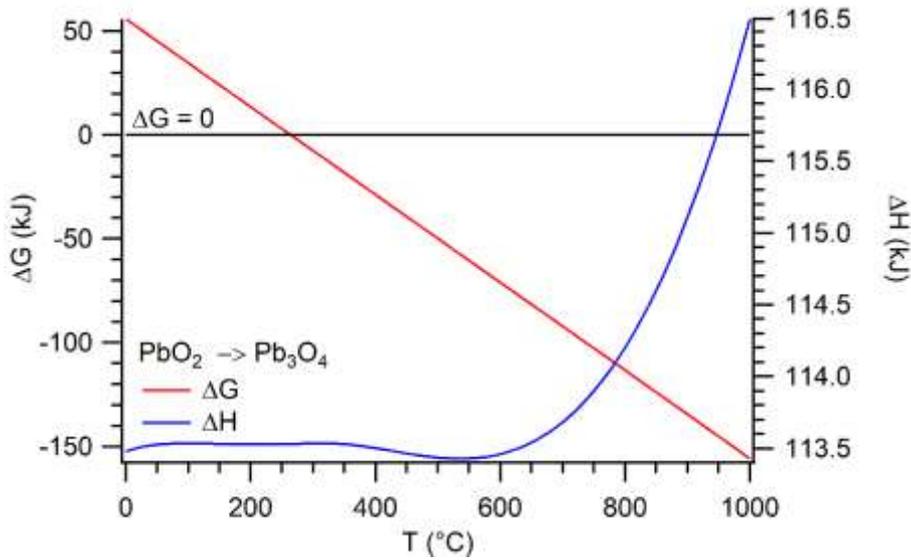

Figure S-4: Gibbs free energy and enthalpy of the $PbO_2$ to $Pb_3O_4$ reaction calculated with the HSC Chemistry software.

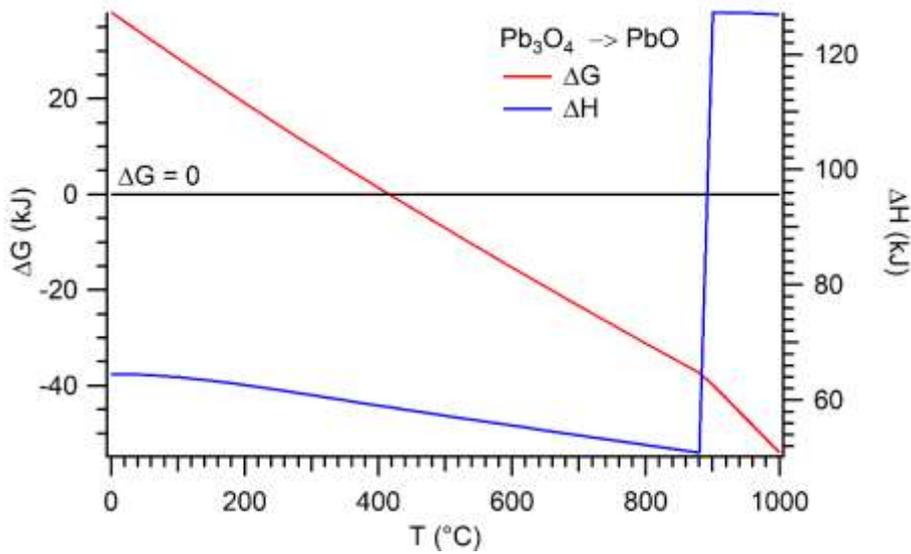

Figure S-5: Gibbs free energy and enthalpy of the $Pb_3O_4$ to $PbO$ reaction calculated with the HSC Chemistry software.

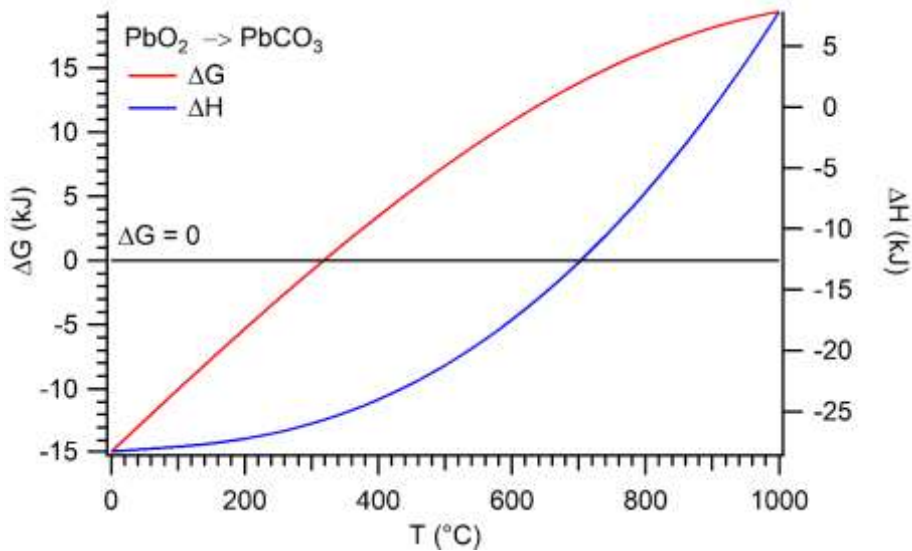

Figure S-6: Gibbs free energy and enthalpy of the $PbO_2$ to $PbCO_3$ reaction calculated with the HSC Chemistry software.

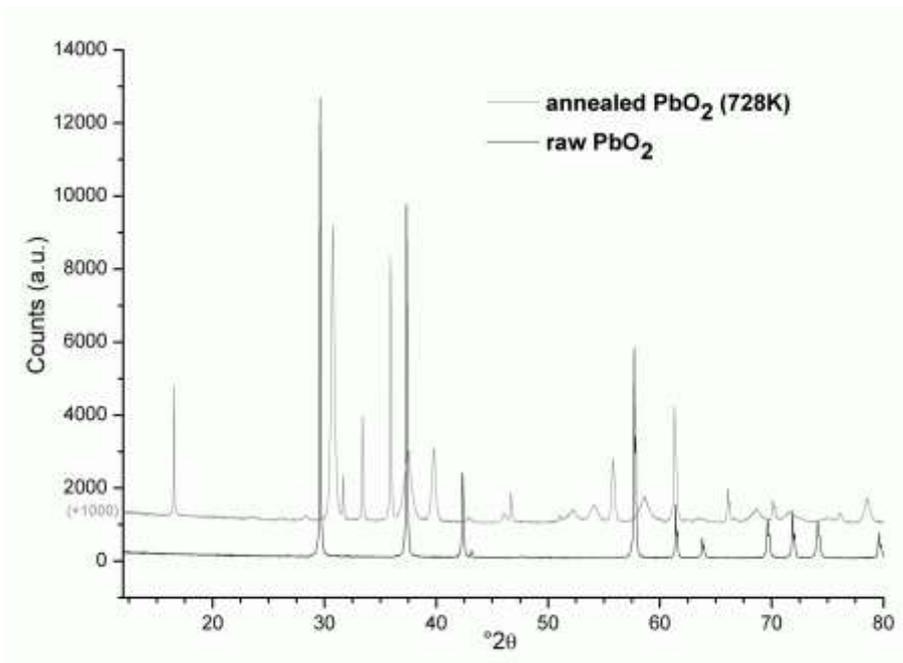

Figure S-7: powder X-ray diffraction patterns of raw plattnerite sample before (β-$PbO_2$) and after calcinations at 728K during 6 hours ($Pb_3O_4$). Initial plattnerite particles show typical coherent domains sizes between 0.8 and 1.5 µm while treated plattnerite particles, consisting of minium crystallites, have notably smaller sizes, [0.1; 1.5] µm (table S-7).

Table S-7: powder X-ray diffraction features of both raw and thermally treated plattnerite powder.

|  | h k l | °2θ | $d_{hkl}$ (Å) | Irel (%) th | $H_{hkl}$ (°2θ) | $<D_{hkl}>$ (µm) |
|---|---|---|---|---|---|---|
| Raw $PbO_2$ | 1 1 0 | 29.61 | 3.500 | 100 | 0.067 | **1.43** |
|  | 1 0 1 | 37.30 | 2.797 | 94.8 | 0.079 | **1.23** |
|  | 2 0 0 | 42.28 | 2.480 | 29.6 | 0.115 | **0.86** |
|  | 2 1 1 | 57.63 | 1.856 | 69.1 | 0.099 | **1.06** |
|  | 2 2 0 | 61.41 | 1.752 | 17.3 | 0.087 | **1.22** |
|  | 1 1 0 | 16.51 | 6.230 | 11.0 | 0.063 | **1.49** |

| | 2 1 1 | 30.69 | 3.380 | 100 | 0.308 | **0.31** |
|---|---|---|---|---|---|---|
| Thermally treated PbO$_2$ (Pb$_3$O$_4$) | 0 0 2 | 31.62 | 3.283 | 5.0 | 0.079 | **1.22** |
| | 2 2 0 | 33.36 | 3.116 | 13.0 | 0.073 | **1.31** |
| | 1 1 2 | 35.87 | 2.905 | 37.7 | 0.085 | **1.13** |
| | 3 1 0 | 37.43 | 2.788 | 48.7 | 0.801 | **0.12** |
| | 2 0 2 | 39.72 | 2.633 | 26.1 | 0.340 | **0.29** |

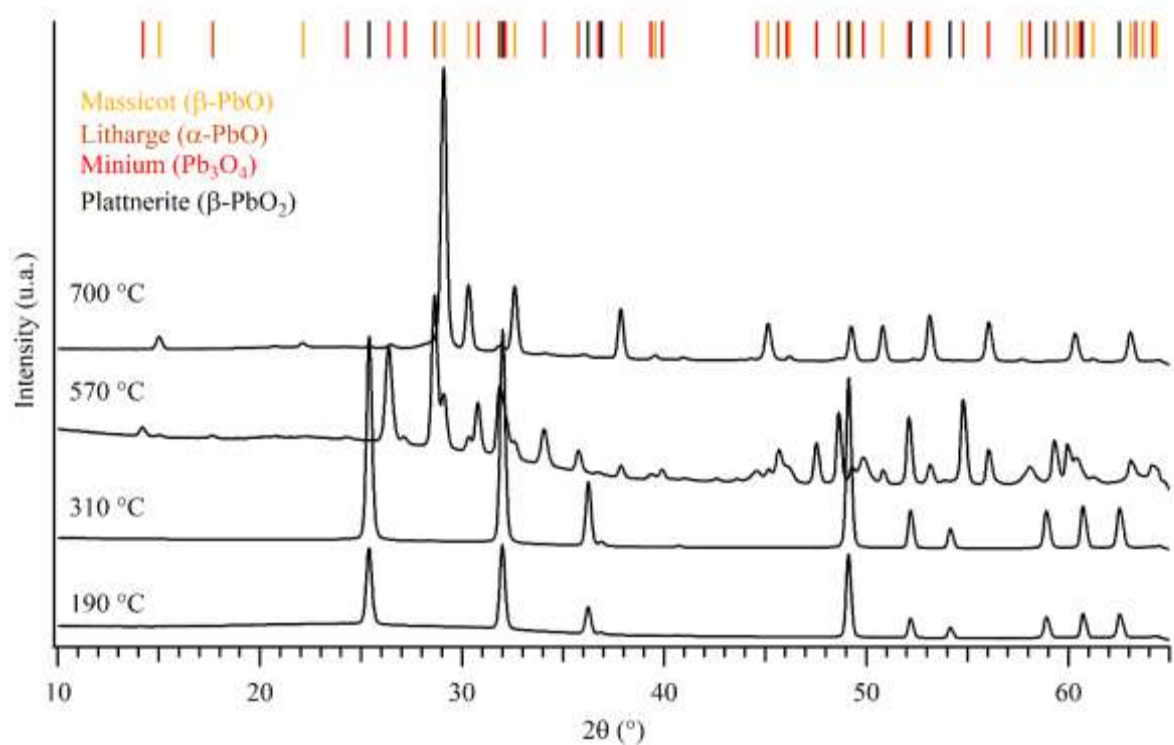

Figure S-8: µ-XRD diagram of the product obtained by TGA at various temperatures.

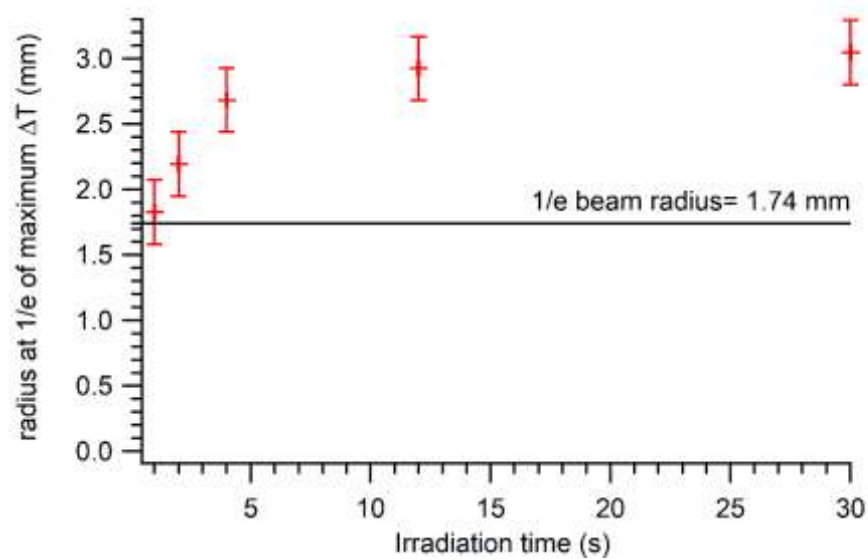

Figure S-9: evolution of the radius at 1/e of maximum ΔT during irradiation described on figure 3a. Error bars on the radius come from the 0.2439 mm step of the IR camera.

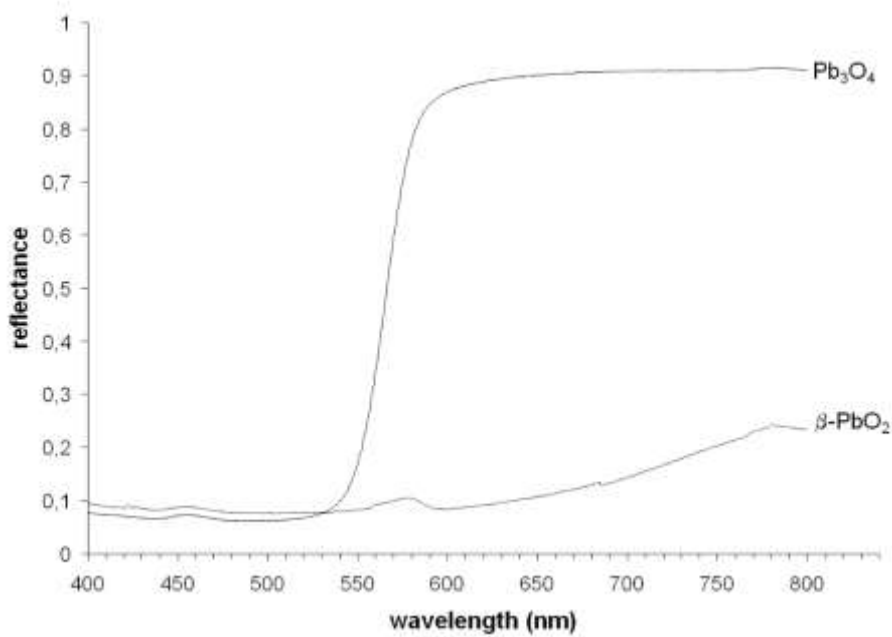

Figure S-10: visible total reflectance spectra of standard minium (Pb$_3$O$_4$) and plattnerite (β-PbO$_2$). These spectra were collected on thick powder layers using a Perkin Elmer Lambda 35 spectrometer equipped with a 60 mm integrating sphere.

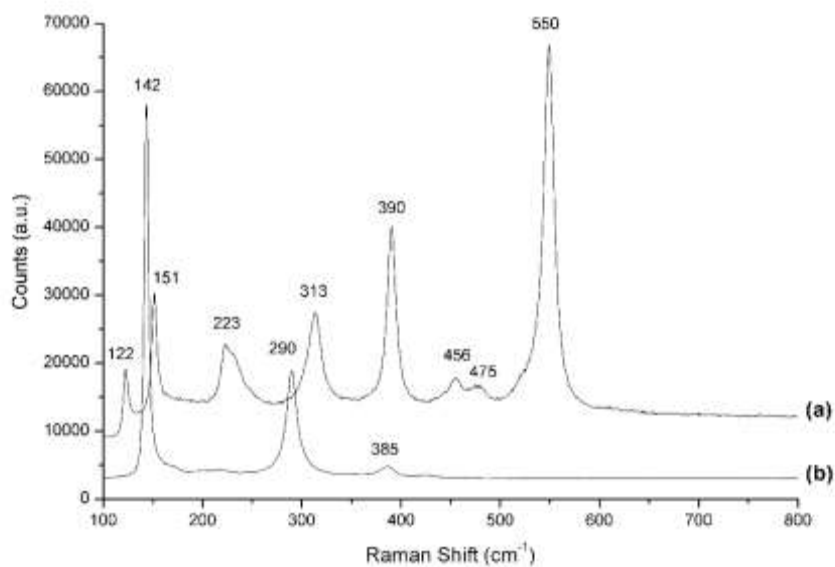

Figure S-11: Raman spectra of raw plattnerite pellet after laser irradiation test using a Gaussian shaped CW Ar$^+$ laser at a 370 kW m$^{-2}$ irradiance. Spectra were acquired in the outer area (a, characteristic bands of minium Pb$_3$O$_4$) and in the central area (b, characteristic bands of massicot β-PbO).

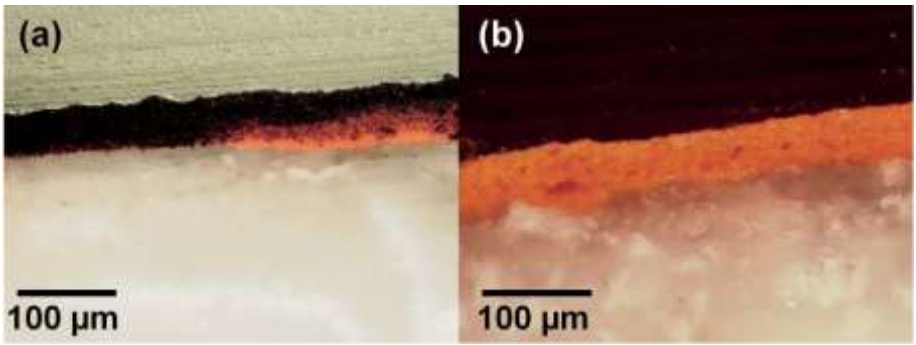

Figure S-12: cross-sections of the naturally aged red lead paint sample showed in figure 6 in (a) the original darkened area; (b) the reconverted red area after CW top-hat Nd:YAG laser irradiation at 1000 kW m$^{-2}$ for 5 seconds.

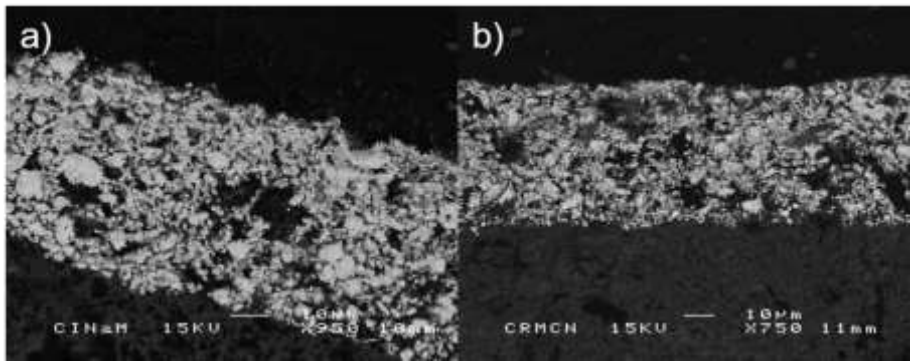

Figure S-13: backscattered electron micrographs of the paint cross-sections showed in figure 6 in (a) the original blackened area, mainly constituted by plattnerite; (b) the reconverted red area constituted by minium grains.

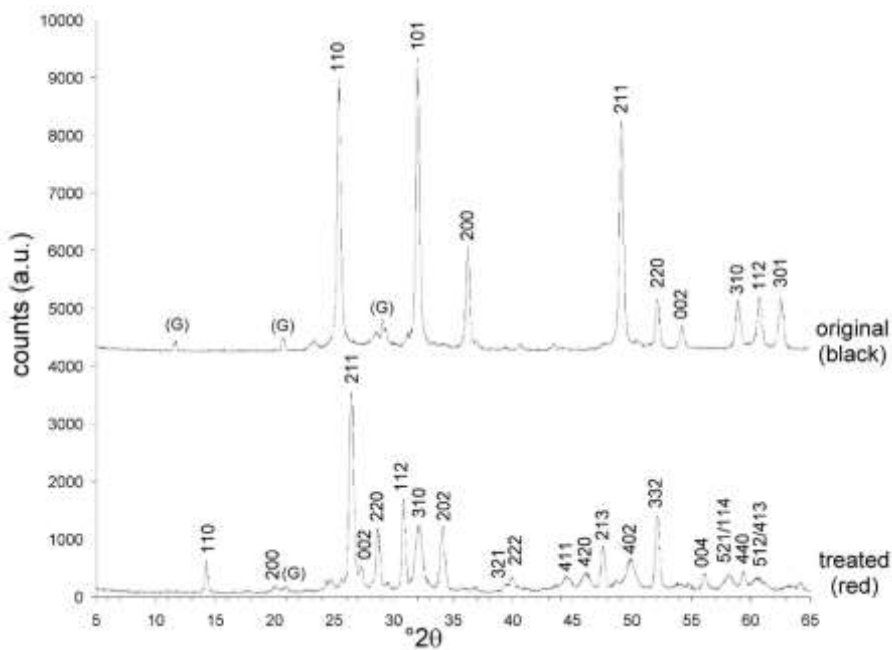

Figure S-14: powder X-ray diffraction patterns of the original black layer (upper) and the reconverted red area (low) of the sample showed in figure 6. Indexed (hkl) reflections of plattnerite ($\beta$-PbO$_2$, upper) and minium (Pb$_3$O$_4$, low) are displayed.

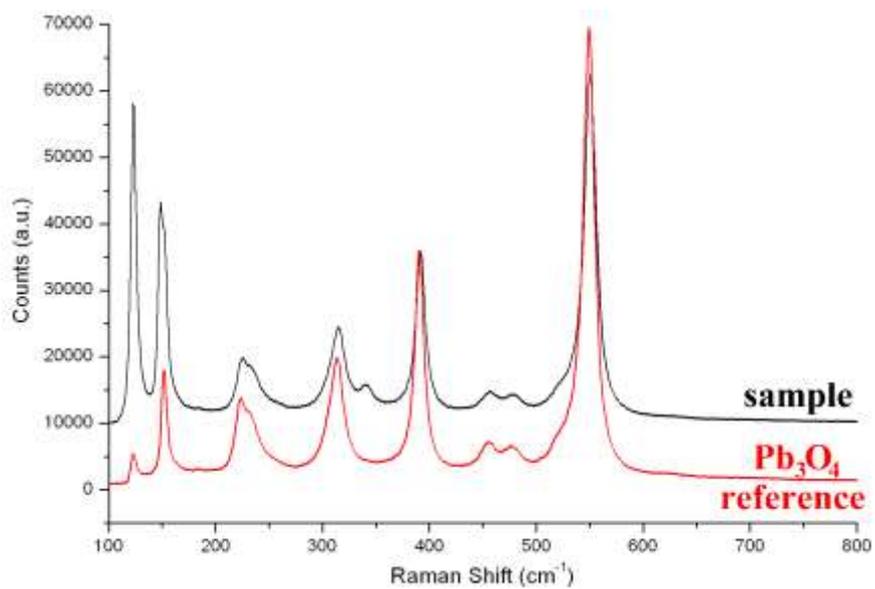

Figure S-15: Raman spectra of a reconverted red area and of pure minium reference.